\documentclass[twocolumn]{aastex631}

\usepackage{amsmath}

\usepackage{color}
\usepackage{url}
\usepackage{hyperref}
\usepackage{xspace}
\usepackage{soul}
\usepackage{hyperref}

\newcommand{\source}{\mbox{Cir~X-1}\xspace}
\newcommand {\nicer}{{NICER}\xspace} 
\newcommand {\ixpe}{{IXPE}\xspace}
\newcommand {\nustar}{{NuSTAR}\xspace}
\newcommand {\flux}{erg\,s$^{-1}$\,cm$^{-2}$\xspace}

\newcommand{\angleseg}{$49\degr\pm8\degr$}
\newcommand{\anglehr}{$67\degr\pm11\degr$}

\begin{document}

\title{X-Ray Polarized View on the Accretion Geometry in the X-Ray Binary Circinus X-1}

\author[0000-0002-9774-0560]{John Rankin}
\affiliation{INAF Istituto di Astrofisica e Planetologia Spaziali, Via del Fosso del Cavaliere 100, 00133 Roma, Italy}

\author[0000-0001-8916-4156]{Fabio La Monaca}
\affiliation{INAF Istituto di Astrofisica e Planetologia Spaziali, Via del Fosso del Cavaliere 100, 00133 Roma, Italy}
\affiliation{Dipartimento di Fisica, Universit\`{a} degli Studi di Roma ``Tor Vergata'', Via della Ricerca Scientifica 1, 00133 Roma, Italy}
\affiliation{Dipartimento di Fisica, Universit\`{a} degli Studi di Roma ``La Sapienza'', Piazzale Aldo Moro 5, 00185 Roma, Italy}

\author[0000-0003-0331-3259]{Alessandro Di Marco}
\affiliation{INAF Istituto di Astrofisica e Planetologia Spaziali, Via del Fosso del Cavaliere 100, 00133 Roma, Italy}

\author[0000-0002-0983-0049]{Juri Poutanen}
\affiliation{Department of Physics and Astronomy,  20014 University of Turku, Finland}

\author[0009-0009-3183-9742]{Anna Bobrikova}
\affiliation{Department of Physics and Astronomy,  20014 University of Turku, Finland}

\author[0000-0002-7502-3173]{Vadim Kravtsov}
\affiliation{Department of Physics and Astronomy, 20014 University of Turku, Finland}

\author[0000-0003-3331-3794]{Fabio Muleri}
\affiliation{INAF Istituto di Astrofisica e Planetologia Spaziali, Via del Fosso del Cavaliere 100, 00133 Roma, Italy}	

\author[0000-0001-7397-8091]{Maura Pilia}
\affiliation{INAF Osservatorio Astronomico di Cagliari, Via della Scienza 5, 09047 Selargius (CA), Italy}

\author[0000-0002-5767-7253]{Alexandra Veledina}
\affiliation{Department of Physics and Astronomy, 20014 University of Turku, Finland}
\affiliation{Nordita, KTH Royal Institute of Technology and Stockholm University, Hannes Alfv{\'e}ns v{\"a}g 12, SE-106 91, Sweden} 

\author[0000-0002-5654-2744]{Rob Fender}
\affiliation{Department of Physics, University of Oxford, Denys Wilkinson Building, Keble Road, Oxford OX1 3RH, UK}

% ...

\begin{abstract}
\source is a neutron star X-ray binary characterized by strong variations in flux during its eccentric $\sim$16.6 days orbit. There are also strong variations in the spectral state, and historically it has shown both atoll and Z state properties.
We observed the source with the Imaging X-ray Polarimetry Explorer during two orbital segments, 6 days apart, for a total of 263~ks.
We find an X-ray polarization degree in these segments of $1.6\%\pm0.3\%$ and $1.4\%\pm0.3\%$ at polarization angles of $37\degr\pm5\degr$ and $-12\degr\pm7\degr$, respectively. 
Thus we observed a rotation of the polarization angle by \angleseg\ along the orbit. 
Because variations of accretion flow, and then of the hardness ratio, are expected during the orbit, we also studied the polarization binned in hardness ratio, and found the polarization angle differing by \anglehr\ between the lowest and highest values of the hardness ratio.
We discuss possible interpretations of this result that could indicate a possible misalignment between the symmetry axes of the accretion disk and the Comptonizing region caused by the misalignment of the neutron star's angular momentum with respect to the orbital one. 
\end{abstract}

\section{Introduction}

X-ray binaries (XRBs) consist of a compact object with a stellar companion orbiting it, from which it accretes matter. Flux and spectral variations in XRBs are thought to correspond to different accretion configurations. The spectrum of each state can be interpreted as a superposition of different components having a different relative flux: typically the accretion disk, emitting a soft nearly thermal spectrum described by a blackbody or a multicolor disk, and a corona of hot plasma, whose electrons up-scatter the low-energy ambient photons, generating a hard X-ray component. In neutron star (NS) X-ray binaries (NS-XRBs), the surface region, where the accreting matter is stopped, also contributes to the total emission. The interfacing region, which is coplanar to the accretion disk, is known as the boundary layer  \citep[BL;][]{Shakura88, Popham01}, while the gas layer at the NS surface, extending up to high latitudes, is known as the spreading layer \citep[SL;][]{inogamov1999,Suleimanov2006,Abolmasov2020}. XRBs with a weakly magnetized NS are classified, according to their tracks on the hard/soft X-ray color diagram, as Z or atoll sources \citep{vanderklis89,hasinger89}. 

\source is a weakly magnetized NS-XRB characterized by an eccentric ($e\sim 0.45$) $\sim$16.5~d orbit 
\citep[see e.g.][]{Kaluzienski1976,Schulz2020}, during which its flux and spectrum change significantly, very different from any other known XRB.
\source has historically been shown to go through all the different states for both Z and atoll sources \citep{2019IAUS..346..125S}. On the basis of its spectral characteristics, it was for a long time suspected to host a black hole, but the discovery of type~I bursts undoubtedly proved that the compact object is a NS \citep{1986MNRAS.221P..27T, 2010ApJ...719L..84L}. During its orbit, the X-ray flux varies by two orders of magnitude, and there are also more irregular decades-long variations \citep{2012A&A...543A..20D}.  

An extended emission from a supernova remnant has been found around \source \citep{2013ApJ...779..171H} with an estimated age of 4600 years, implying that the source is the youngest known XRBs. The young age is consistent both with the eccentricity of the orbit and with its irregular variations.
Another characteristic that sets \source apart from other XRBs is the presence of both radio and --- unique case identified so far for NS-XRBs --- X-ray jets, indicating the ejection of matter at relativistic speeds. The X-ray jets are clearly visible on both sides of the receding and the approaching radio jet  \citep{2007ApJ...663L..93H, 2009MNRAS.397L...1S}. Their presence, observed by Chandra either as a Doppler shift of emission lines or by directly imaging a diffuse and elongated emission, is interesting for comparing this system with black holes, showing that jets can be produced despite the shallower gravitational potential of NSs \citep{2004Natur.427..222F}. The orientation of the radio jets has been reported to change with time either because of the precession of the regions from which it is emitted, which is well-accepted that they must be close to the compact object, or because of the interaction with the interstellar matter \citep{2019MNRAS.484.1672C}. 

Several models have been proposed to account for the peculiar orbital and state variations of \source. A dip in the light curve -- followed by a flaring phase -- is seen every orbit. This dip could be caused by a cold absorber; however, this would work only for high inclinations \citep{2012A&A...543A..20D}. 
According to \citet{1999MNRAS.308..415J} the eccentric orbit causes orbital variations in the mass accretion rate, producing the modulation in the X-ray luminosity. \citet{2019IAUS..346..125S} suggested that the companion is a massive supergiant of a Be type, which would imply that \source is a high-mass Be XRB. The available observations until now have not been sufficient to discriminate among the different models.

\begin{figure*}[!ht]
    \centering
\includegraphics[width=0.87\linewidth]{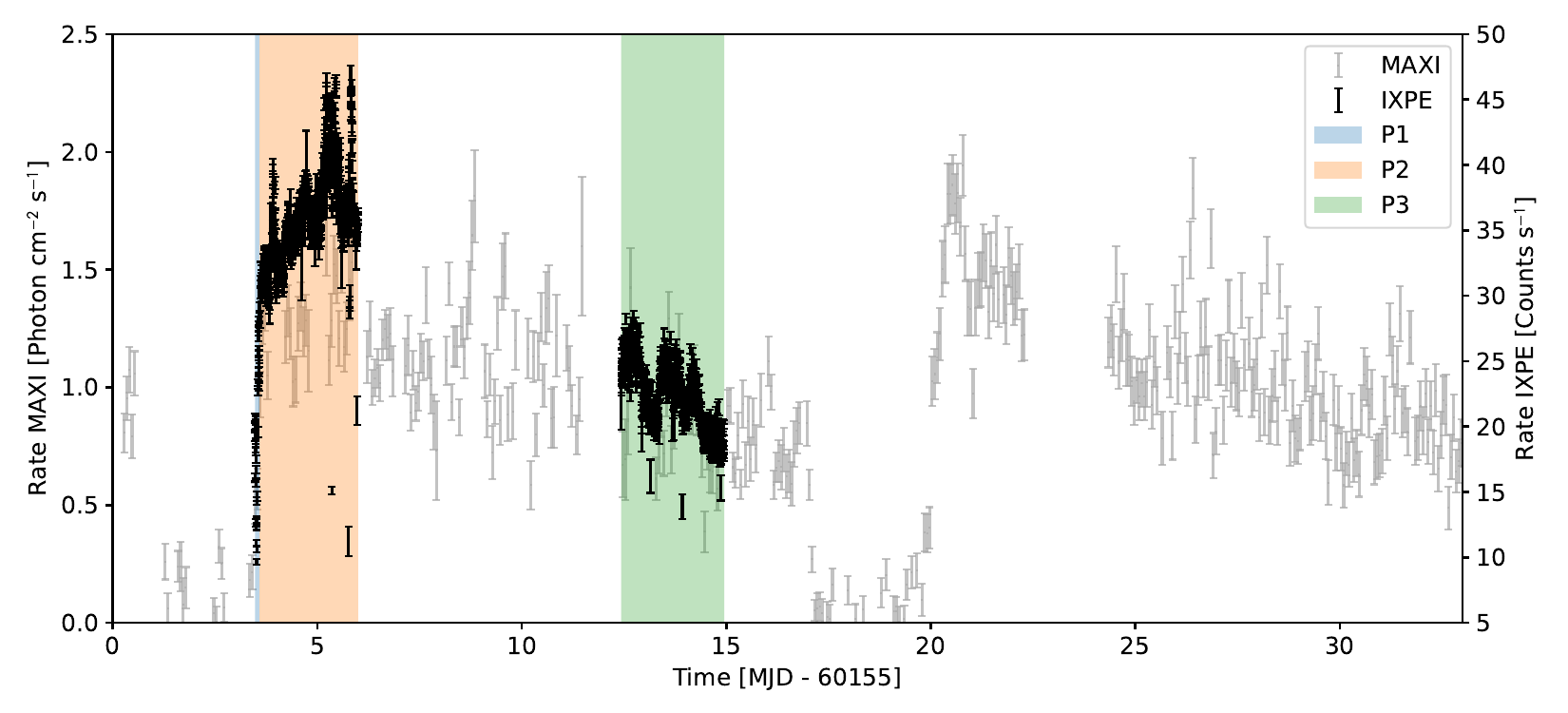}
\includegraphics[width=0.87\linewidth]{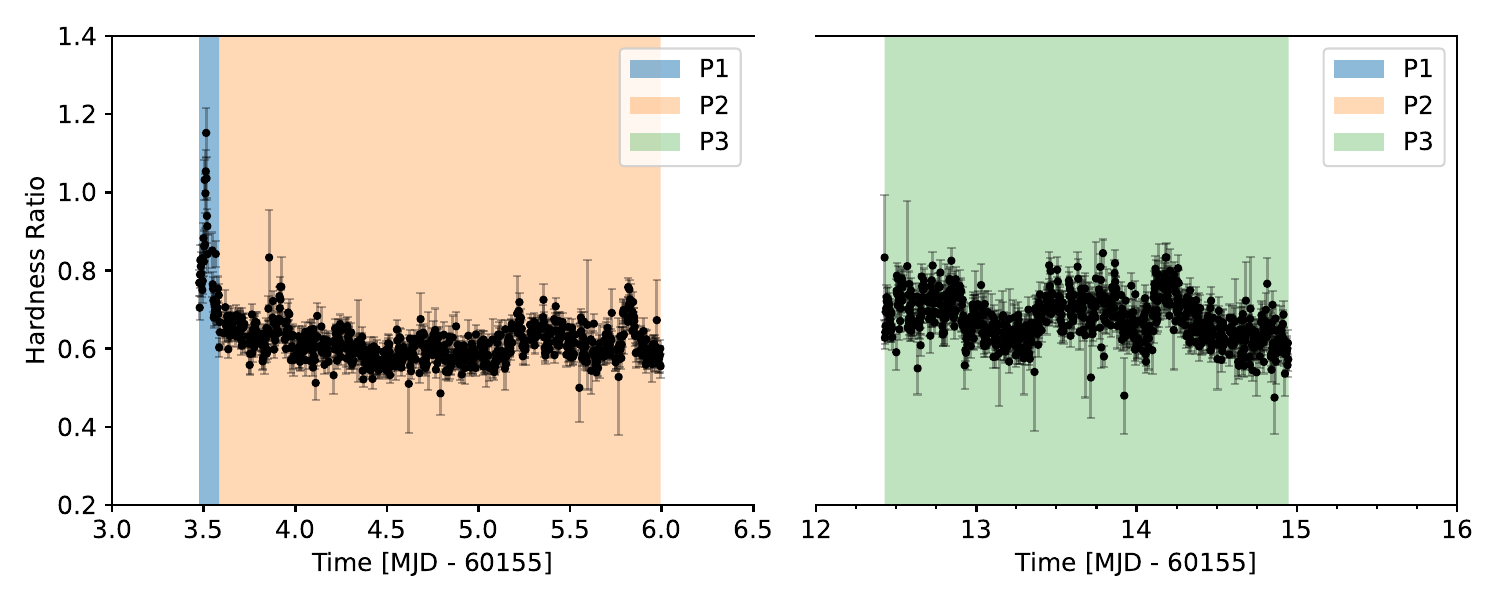}
\caption{Evolution of the X-ray properties of \source. (Top) Rate obtained by MAXI (2--20~keV) and IXPE over two orbits binned in 180\,s time bins. 
(Bottom) Hardness ratio obtained from IXPE (Eq. \ref{eq:HRixpe}) during the observations binned in 500\,s time bins. There is a clear change in state from the hard to the soft at the beginning of the IXPE observation; based on this, we divide the analysis in the 3 phase intervals indicated in the plots by the colored regions.}
\label{fig:rate-HR-trends}
\end{figure*}

We report the first polarimetric study of \source using IXPE, measuring the new observables of polarization degree (PD), and polarization angle (PA). With X-ray polarimetry we can discern, from different polarization signatures, the different emission mechanisms, and the geometry of the regions closer to the compact object. In the absence of relativistic effects the PA is expected to be either parallel or perpendicular to the main geometrical axis of the component; as a consequence, if two components (such as disk and Comptonized region) are aligned, we expect their PAs to be either the same or orthogonal. If this is not the case, it can indicate that there is a geometrical misalignment between them, or that relativistic effects rotate the polarization plane \citep[e.g.,][]{Connors1977,Connors1980,Dovciak2004,Loktev2020,Loktev2022}.

\section{Observations}
\subsection{IXPE}

The Imaging X-ray Polarimetry Explorer \citep[IXPE;][]{Weisskopf2022, 2021AJ....162..208S} is the first observatory combining detectors sensitive to X-ray polarization in the 2--8~keV energy band with X-ray optics. This mission, a collaboration between NASA and the Italian Space Agency (ASI), consists of three X-ray polarization sensitive Gas Pixel Detectors  \citep[GPD;][]{2001Natur.411..662C, 2006NIMPA.566..552B, 2007NIMPA.579..853B, 2021APh...13302628B} at the focus of three grazing incidence optics. Other than detecting polarization, IXPE simultaneously detects the energy, time of arrival and celestial position of each X-ray detected.

IXPE observed \source in two different pointings (2023-08-02T11:24 to 2023-08-04T23:52, and 2023-08-11T10:17 to 2023-08-13T22:46, joint in the ObsID 02002699, see Table \ref{tab:obs_ids}), to cover two different parts of the orbit, for a net total exposure time of 263~ks. \nicer and \nustar observations were performed  to simultaneously partially cover the \ixpe observation. IXPE data are reduced and corrected by the standard pipeline running at the Science Operations Center in NASA/MSFC, and were downloaded from the IXPE public archive at HEASARC.\footnote{https://heasarc.gsfc.nasa.gov/docs/ixpe/archive/}
In the following analysis, event-by-event Stokes parameters are calculated following an unweighted approach \citet{2015APh....68...45K} and  \citet{2022AJ....163..170D}, and computed using \textsc{ixpeobsssim 30.6.3} \citep{2022SoftX..1901194B}; they are provided to the user in a reference frame projected on the sky. We selected the source in a circular region of radius 90\arcsec\ centered on the source.
Because of the high brightness of this source, the background is negligible \citep[see][]{2023AJ....165..143D}.

\begin{figure*}
\centering
\includegraphics[width=1\columnwidth]{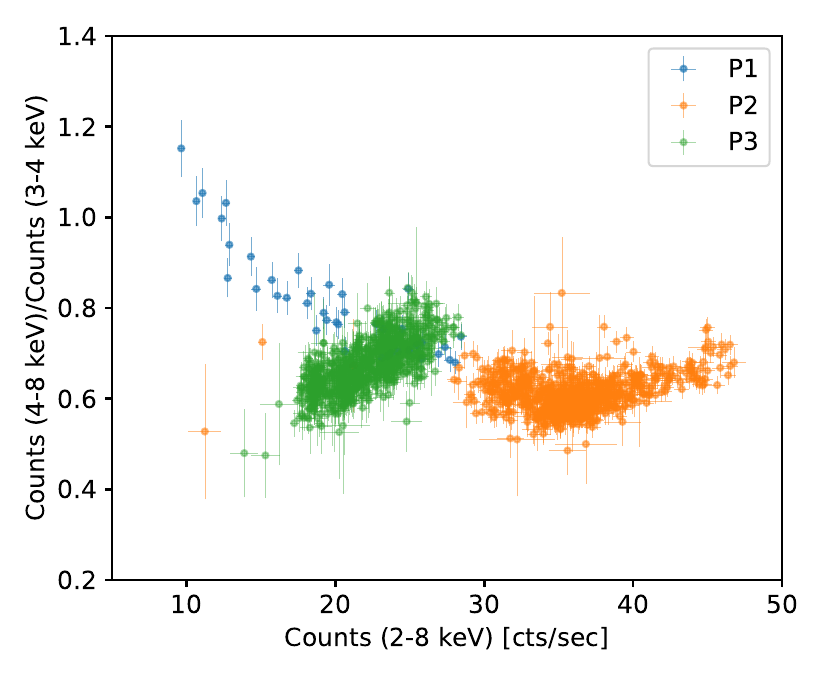}
\includegraphics[width=1\columnwidth]{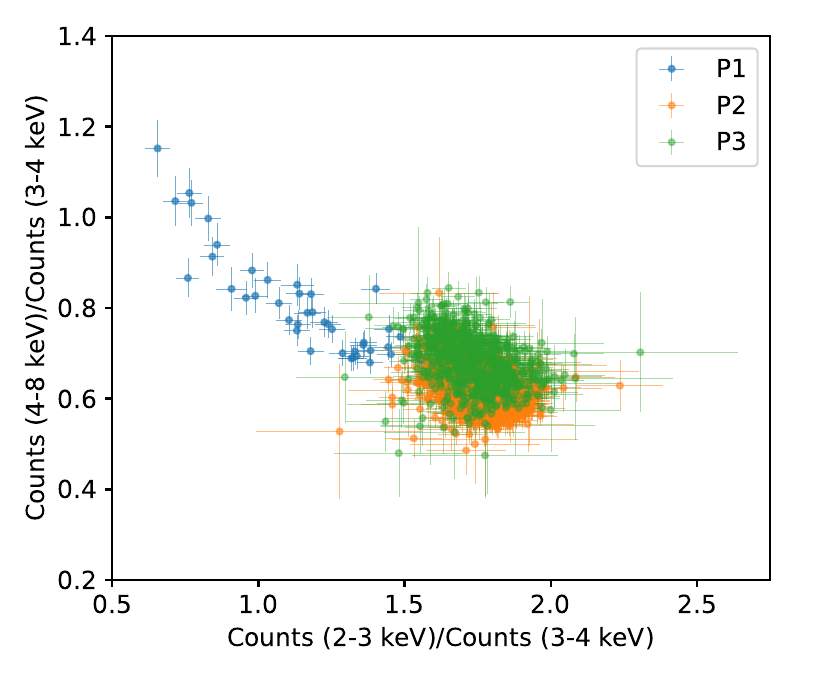}
\caption{Hardness-intensity (left) and color-color (right) diagrams during the IXPE observations, binned in time bins of 180\,s, with color highlighted the three phase intervals chosen for the subsequent analysis.  
}
\label{fig:hid-cc}
\end{figure*}

\begin{deluxetable*}{lcccc}
\tablecaption{Observational data used in this paper, reporting for each mission the observation IDs, the livetime and the start and end times of each observation. \label{tab:obs_ids}}
\tablehead{
 \colhead{Mission} & \colhead{Obs ID} & \colhead{Livetime [s]} & \colhead{Start Time} & \colhead{End Time }
}
\startdata
IXPE &	2002699 &	263000 & 2023-08-02T11:24 & 2023-08-04T23:52 \\
 & & & 2023-08-11T10:17 & 2023-08-13T22:46 \\
NICER &	6689030104 &	 305 & 2023-08-05T00:22 & 2023-08-05T00:35 \\
 &	6689030203 &	4325 & 2023-08-13T00:27 & 2023-08-13T20:57 \\
NuSTAR &	30902037002 &	12000 & 2023-08-04T21:51 & 2023-08-05T09:01 \\
 &	30902037004 &	15000 & 2023-08-12T14:31 & 2023-08-13T01:26 \\
\enddata
\end{deluxetable*}

The top panel of Figure~\ref{fig:rate-HR-trends} shows the IXPE light curve during these observations --- overlaid with data from the Monitor of All-sky X-ray Image \citep[MAXI;][]{2009PASJ...61..999M} telescope. MAXI is mounted on-board the International Space Station, and monitors X-ray sources continuously; therefore its light curve allows to study the flux variations in \source over its entire orbit, even outside the IXPE observation.

To verify possible changes in the accretion flow of \source, we study its flux, and its Hardness-Ratio (HR) variations over the IXPE observing time. The bottom panel of Figure~\ref{fig:rate-HR-trends} shows the IXPE HR, defined as
\begin{equation}
HR_\text{IXPE} =  \frac{\text{Counts\,[4--8\,keV]}}{\text{Counts\,[3--4\,keV]}} , \label{eq:HRixpe}
\end{equation}
The same plots show a division of the overall observations in three phase intervals, which we will use in the following to study the polarization along the orbit of \source: P1 (phase from 0.21 to 0.22), P2 (phase from 0.22 to 0.36) and P3 (phase from 0.75 to 0.90). We clearly see during the first observation a transition from a low flux --- hard --- state, to a high flux --- soft --- state. The IXPE observation starts just when the source is coming out from the dip, as shown by the MAXI light curve, so that the low-hard state corresponds to this part of the orbit. 

Figure~\ref{fig:hid-cc} (left) shows the hardness-intensity diagram (HID) for the three phase intervals obtained from the IXPE data. We clearly see a variation in hardness-intensity between the low-hard and high-soft state when moving from the first to the second phase interval. We also see that the HR is on average slightly larger in the third phase interval, when the flux was lower, compared to the second phase interval.  
The same effect is also seen in the color-color diagram (right panel of Figure~\ref{fig:hid-cc}), which shows the evolution of the source in two colors defined for the low- and high-energy bands.

\citet{2023ApJ...958...52T} studied \source for an extended period and divided (figure 2 of their paper) the orbit in different phases: a dip phase, where the X-ray flux is low due to strong absorption, whose end corresponds to P1 in this paper; a flaring phase, with rapid changes, corresponding to P2 in this paper; and a stable phase, with a gradual decrease in X-ray flux, corresponding to P3 in this paper.

\subsection{NICER}

The Neutron Star Interior Composition Explorer \citep[NICER;][]{2016SPIE.9905E..1HG}, mounted on-board the International Space Station, observed \source during part of the IXPE observations. Obsid 6689030104 --- right at the end of the first IXPE observation --- and obsid 6689030203 --- during the second IXPE observation (see Table \ref{tab:obs_ids}) --- were used to study the spectral components of \source (see Section \ref{sec:spec}).
\nicer, consists of 56 co-aligned concentrator X-ray optics, each with a silicon drift detector at its focus, and, although it does not have imaging capabilities, it has a large collecting area in the energy interval of 0.2--12 keV. These observations were obtained in the framework of the GO Cycle 5 (proposal 6189); data were processed with the NICER Data Analysis Software v010a released on 2022 December 16 provided under \textsc{HEASoft} v\,6.31.1 with the CALDB version released on 2022 October 30.

\begin{figure*}
\centering
\includegraphics[width=0.49\linewidth]{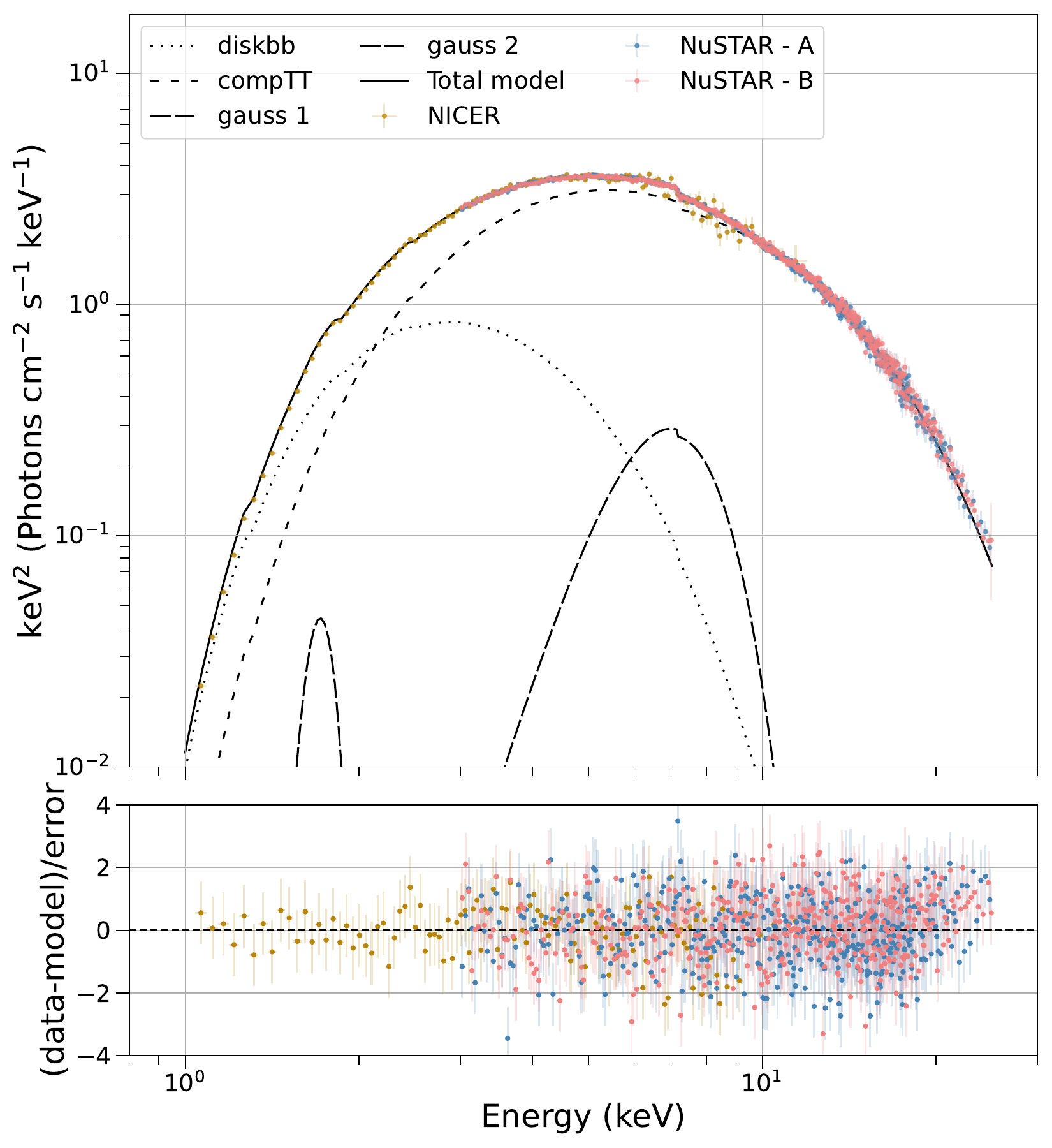}
\includegraphics[width=0.49\linewidth]{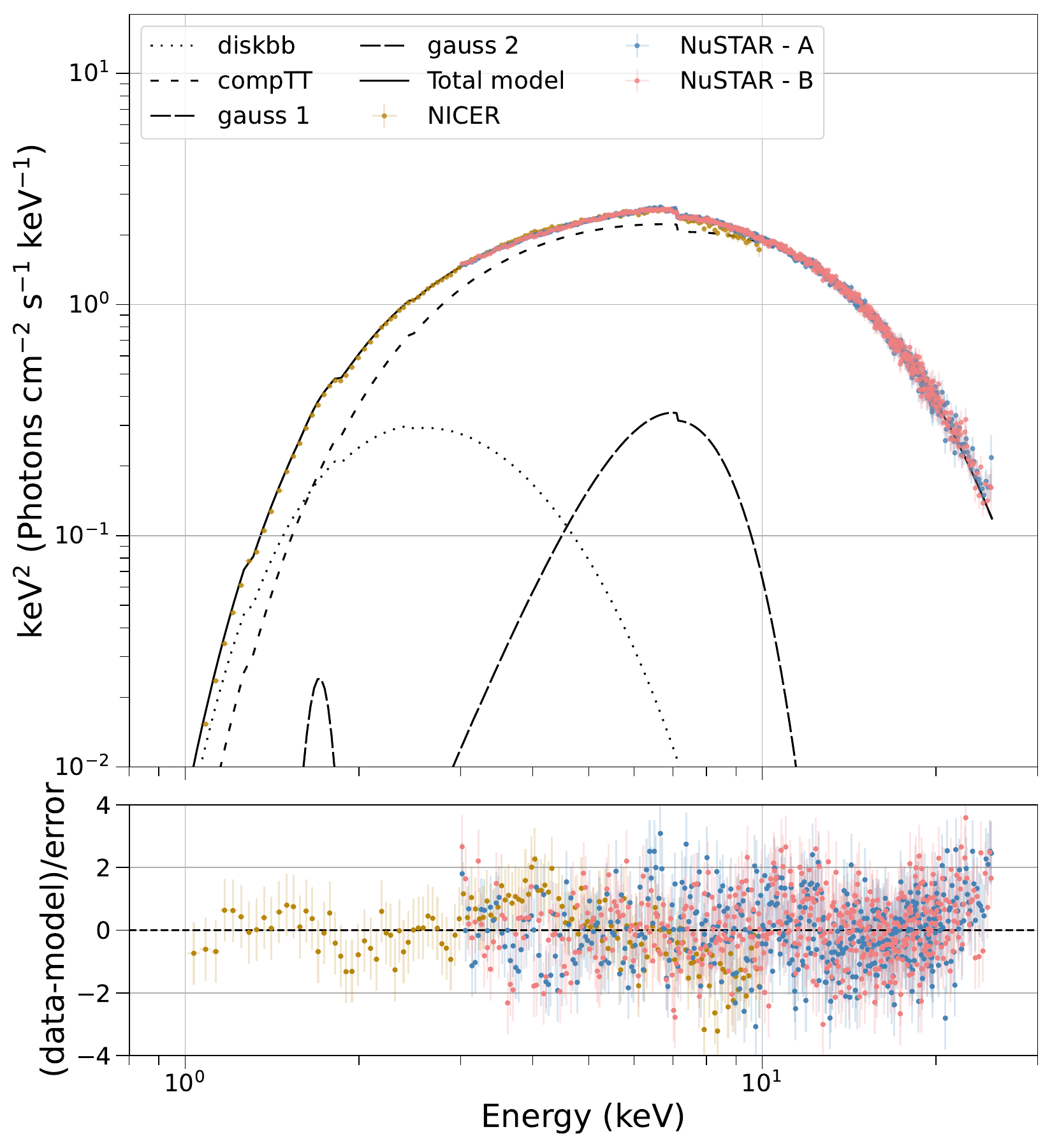}
\caption{Spectral energy distribution of Cir X-1 during P2 (left) and P3 (right) in $EF_E$ representation using   \nicer and \nustar (FPMA and FPMB) data and showing the different model components.}
\label{fig:spec_fit_NiNu}
\end{figure*}

\subsection{NuSTAR}

The Nuclear Spectroscopic Telescope Array \citep[NuSTAR;][]{Harrison2013} consists of two focal plane modules (FPMA and FPMB), providing broadband X-ray imaging, spectroscopy, and timing in the energy range of 3--79 keV with an angular resolution of 18\arcsec\ (FWHM) and spectral resolution of 400~eV (FWHM) at 10~keV, and it is the only observatory employing multi-layer X-ray optics capable of focusing hard X-rays. We used the \source observations at the end of the first IXPE observation (ObsID 30902037002)  and during the second one (ObsID 30902037004, see Table \ref{tab:obs_ids}), performed in the framework of GO cycle 9 (proposal 9212).

\nustar data were processed by using the standard Data Analysis Software (\textsc{nustardas} v2.1.2 from 2022 Feb 12) provided under \textsc{HEASoft} v\,6.31.1 with the CALDB version released on 2023 April 4. The  source was selected from a circular 150\arcsec\ radius region centered on the source position; the background was extracted in a similar region, but in a position of the field of view out of the source.

\section{Spectroscopic Analysis \label{sec:spec}}

Aiming to constrain the spectral model, and understand the different components, we analyzed \nicer (in 1--10\,keV) and \nustar (in 3--25\,keV) data: the Obsid are these reported above, and were selected to overlap and have a short duration, so that there would be no HR variations. 
Previous spectral fits, such as in the broad-band BeppoSAX spectra \citep{Iaria2002,Iaria2005},  reported the presence of two components: a blackbody disk and a Comptonization component. However the temperature of one of the components --- the disk --- has been reported to be low  \citep[$\sim$0.5~keV,][]{Iaria2008} and so the disk is not expected to contribute significantly to the IXPE energy band. 

\begin{deluxetable*}{crcc}
\tablecaption{Best-fit parameters of the spectral model \texttt{tbfeo*(diskbb+comptt+gauss+gauss)} applied to the simultaneous data from NICER and \nustar. 
}
\label{tab:spec_Ninu}
\tablehead{
\colhead{Model} &
\colhead{Parameter} &
\colhead{P2}  & 
\colhead{P3}  
}
\startdata
    \texttt{tbfeo} & $N_{\rm H}$ ($10^{22}$ cm$^{-2}$) & $3.0^{+0.5}_{-0.8}$ & $3.09^{+0.01}_{-0.31}$  \\
     & Fe ($10^{22}$ cm$^{-2}$) & $<1.6$ & $<0.3$  \\
     & O ($10^{22}$ cm$^{-2}$) & $2.6_{-0.6}^{+1.1}$ & $2.7_{-0.3}^{+0.2}$  \\
    \hline
    \texttt{diskbb} & $kT_{\rm in}$ (keV) & $0.85$\tablenotemark{a} & $0.7$\tablenotemark{a} \\
    & \text{norm} ($[R_{\rm in}/D_{10}]^2\cos\theta$) & $417^{+32}_{-35}$ & $360^{+12}_{-2}$ \\
    \hline
    \texttt{comptt} & $kT_0$ (keV) & $1.00\pm0.03$ & $0.74\pm0.01$ \\
    & $kT_{\rm e}$ (keV) & $2.56^{+0.03}_{-0.02}$ & $2.539^{+0.006}_{-0.003}$ \\
    & $\tau$ & $4.93^{+0.09}_{-0.12}$ & $7.22^{+0.08}_{-0.02}$ \\
    & \text{norm} & $0.68\pm0.02$ & $0.50^{+0.01}_{-0.05}$ \\
\hline
    \texttt{gauss} & $E_{\rm line}$ (keV) & $6.5\pm0.1$ & $6.04^{+0.01}_{-0.22}$ \\
    & $\sigma$ (keV) & $1.27_{-0.08}^{+0.07}$ & $1.75^{+0.01}_{-0.05}$ \\
    & \text{norm} (photon~s$^{-1}$~cm$^{-2}$) & $0.017\pm0.003$ & $0.034\pm0.002$ \\
    \hline
    \texttt{gauss} & $E_{\rm line}$ (keV) & $1.7$\tablenotemark{a} & $1.7$\tablenotemark{a}  \\
    & $\sigma$ (keV) & $0.08\pm 0.03$ & $0.08$\tablenotemark{a} \\
    & \text{norm} (photon~s$^{-1}$~cm$^{-2}$) & $0.009\pm0.0003$ & $0.005\pm0.001$ \\
    \hline
 & $\chi^2$/dof 
    & 1271/1222 = 1.04 & 1398/1226 = 1.14 \\
    \hline
    \multicolumn{4}{c}{Photon flux in 2--8\,keV} \\
    & $F_{\rm tot}$ ($10^{-8}$ \flux) & $2.09$  & $1.53$ \\
   & $F_{\rm diskbb}/F_{\rm tot}$ & 0.18 & 0.09 \\
   & $F_{\rm comptb}/F_{\rm tot}$ & 0.79 & 0.85 \\
   & $F_{\rm gauss}/F_{\rm tot}$  & 0.03 & 0.06 \\
\enddata
\tablecomments{Errors are reported at 68\% CL.}
\tablenotetext{a}{Frozen.}
\end{deluxetable*}

We attempted to fit the continuum with the two components reported in literature: \texttt{diskbb} \citep{1984PASJ...36..741M, 1986ApJ...308..635M} associated with the disk or NS surface, and \texttt{comptt} \citep{1994ApJ...434..570T} associated to the Comptonization in the BL/SL. We saw an excess in the residuals around 6~keV --- associated with a broad iron line due to reflection from the disk --- and around 1.7 keV --- associated with a silicon line, suspected to be an instrumental \nicer feature due to an incorrect calibration of the response, which becomes visible at the high flux observed from \source.  To estimate the absorption from the interstellar medium, we set the abundances at the \texttt{wilm} values \citep{2000ApJ...542..914W}; we started with a \texttt{tbabs} model, but found there were residuals better taken into account using \texttt{tbfeo}.  The resulting model is written in \textsc{xspec} as \texttt{tbfeo*(diskbb+comptt+gauss+gauss)}. 

We started by leaving free all the parameters, but found the temperature of the disk to be degenerate with its norm. We then fixed the disk temperature to a value minimizing the $\chi^2$ and left the norm free in the subsequent fit: the resulting best-fit models are shown in Figure \ref{fig:spec_fit_NiNu} and the parameters are given in Table \ref{tab:spec_Ninu}. 
We see that the flux during P3 is lower than in P2, and the spectrum is harder. This is also reflected in a significantly higher optical depth $\tau_{\rm p}$ during P3, while the electron  temperature is nearly the same. 
We left a free constant between \nicer and \nustar, and also between the different \nustar modules: the two \nustar modules, normalized to \nicer, have values 1.16 and 1.14 in P2, and 1.39 and 1.37 in P3.
Since \nustar shows calibration uncertainties \citep{madsen22}, we left free a gain offset in the fit. No systematic errors were added to the fits (a known systematic error $<$1.5\% is already applied to \nicer data by the \nicer pipeline \texttt{nicerl3-spect}).
From the relative fluxes of the components in the nominal IXPE energy band of 2--8~keV we see that, even if both components are present, the dominant component is \texttt{comptt}. 
One of the Gaussians is an instrumental \nicer feature, while the other is not visible due to a low \ixpe energy resolution with respect to \nicer and \nustar.
Therefore in the following spectropolarimetric analysis of the \ixpe  data we only considered the  \texttt{comptt} component.

\begin{deluxetable*}{lcccccc}
    \tablecaption{Best-fit parameters obtained by fitting IXPE data with the \textsc{xspec} model  \texttt{const*tbabs*comptt}. \label{tab:xspec_I}}
    \tablehead{
    \colhead{} & \colhead{$\chi^2$/dof} & $N_{\rm H}$ ($10^{22}$ cm$^{-2}$) (fixed) & 
 \colhead{$T_0$ (keV)} & \colhead{$kT_{\rm e}$ (keV)} & \colhead{$\tau_{\rm p}$} & \colhead{norm}
    }
    \startdata 
P2 &  417/435 & 2.77 & 0.50$_{-0.04}^{+0.02}$ &  1.57$_{-0.04}^{+0.03}$ &   11.5$\pm$0.4 &  1.71$_{-0.04}^{+0.05}$ \\
P3 &  446/435 & 2.65 & 0.59$_{-0.03}^{+0.04}$ &  2.15$_{-0.09}^{+0.11}$ &    9.4$\pm$0.4 &  0.77$_{-0.04}^{+0.03}$ \\
\hline 
HR1 & 404/429 & 2.7 & 0.62$_{-0.08}^{+0.02}$ &  1.72$_{-0.16}^{+0.07}$ &    9.1$_{-0.4}^{+1.4}$ &  1.5$_{-0.1}^{+0.2}$ \\
HR2 & 397/435 & 2.7 & 0.50$_{-0.02}^{+0.01}$ &  1.68$_{-0.03}^{+0.02}$ &   11.4$_{-0.1}^{+0.3}$ &  1.21$_{-0.01}^{+0.02}$ \\
HR3 & 398/433 & 2.7 & 0.61$_{-0.05}^{+0.03}$ &  2.1$_{-0.1}^{+0.2}$ &   10.1$_{-0.5}^{+0.9}$ &  0.88$_{-0.04}^{+0.02}$ \\
\enddata
\tablecomments{
Uncertainties are at 68\% confidence level.}
\end{deluxetable*}

\section{Polarization Analysis}

We studied the polarization of \source in the three different phase and HR intervals. We first performed a study independent of any spectral model using the \texttt{pcube} algorithm of the \textsc{ixpeobsssim} software \citep{2022SoftX..1901194B}. Then we also studied the polarization applying the spectro-polarimetric analysis in \textsc{xspec} \citep{Arnaud1996} by fitting the $I$, $Q$, and $U$ spectra. Following the spectral models reported in literature \citep[see e.g.,][]{2012A&A...543A..20D}, we describe the $I$ spectrum with the model: \texttt{tbabs*comptt}, setting the abundances to the values of \cite{2000ApJ...542..914W}, for the \texttt{comptt} model the geometry assumed is a disk \citep{10.1063/1.45591}, the hydrogen column density is fixed at the values found in the fit from the previous section for P2 and P3, and for the intermediate value for the HR bins. We did not fit P1 data since it is too short, has too few counts for polarimetric studies, and it has no contemporaneous \nustar data (however we still studied this data using \texttt{pcube}).  
To take into account IXPE calibration uncertainties \citep{2022SPIE12181E..1CD, 10.1117/12.2677264}, we left free the gain slope and offset, obtaining values of the order respectively of 95\% for the slope, and in the 0.005 to 0.2~keV range for the offset.
The results for the spectral modeling in the two phase intervals and in the three HR intervals are reported in Table~\ref{tab:xspec_I}. 
For the spectro-polarimetric analysis, we use the best-fit spectral fits from Table~\ref{tab:xspec_I}) and then 
fit $Q$ and $U$ using model \texttt{polconst} in \textsc{xspec} (see Table~\ref{tab:xspec_IQU} for the obtained results). 
We also attempted fitting with the \texttt{pollin} model, but this gave no improvement over \texttt{polconst}. We considered fitting with a double \texttt{polconst}  model ( \texttt{tbabs*(comptt*polconst + diskbb*polconst)}). However in this case, due to the low flux of diskbb, not all bins under consideration provide an acceptable fit with all parameters constrained and we are not able to separate the two components.

\begin{figure*} 
\centering
\includegraphics[width=0.50\linewidth]{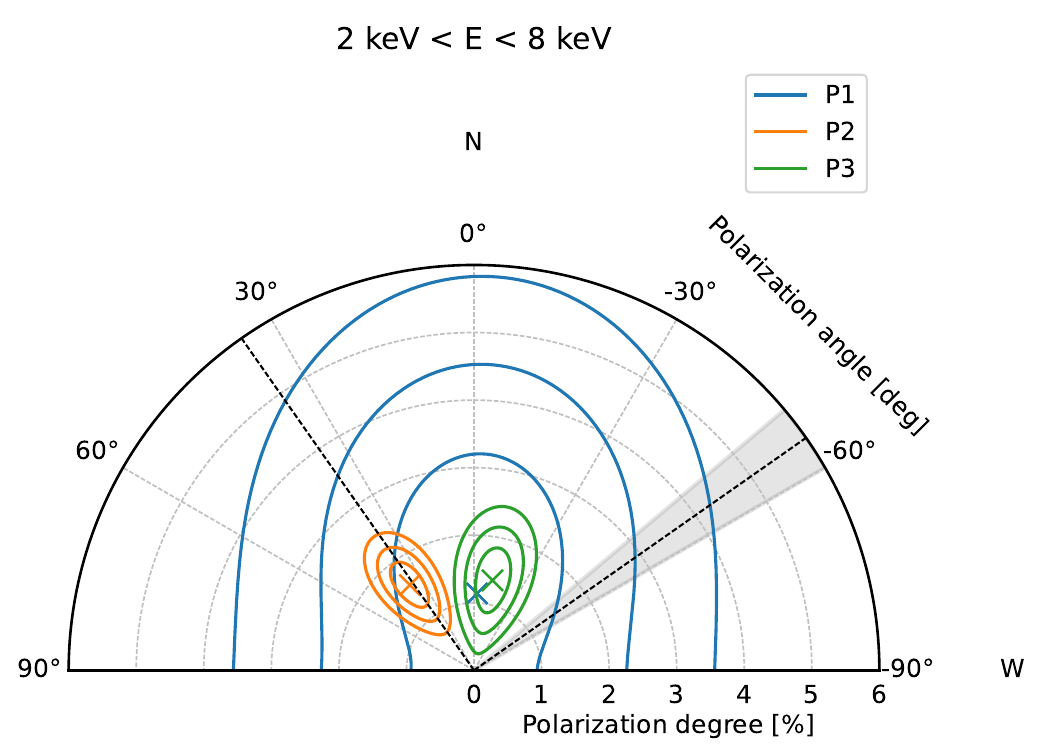}
\includegraphics[width=0.95\linewidth]{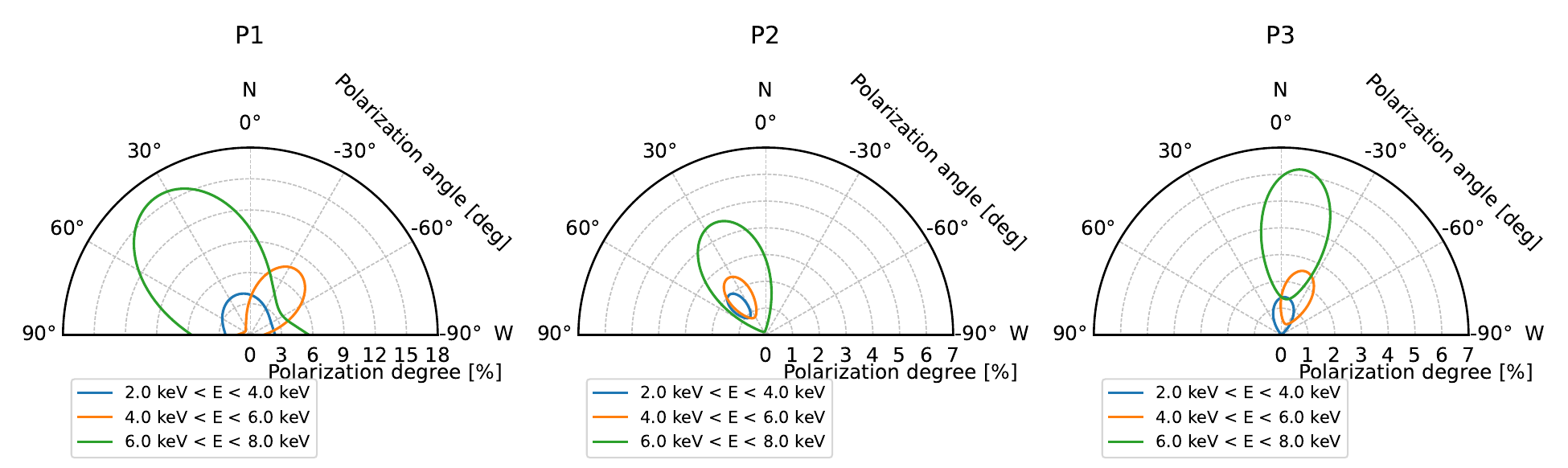}
\caption{Polar plot of polarization, computed using the \texttt{pcube} algorithm from the \textsc{ixpeobsssim} software \citep{2022SoftX..1901194B}, for the three phase intervals defined in Figure~\ref{fig:rate-HR-trends}. 
(Top) Polarization in the entire IXPE energy band is reported. The shaded region indicates the direction of the jet (see discussion), and the black lines indicate this direction and its orthogonal direction. Contours are reported at the 68\%, 95\%, and 99\% confidence levels. 
(Bottom) polarization in different energy bands for the three phase intervals, with contours showing 90\% confidence level.
}
\label{fig:pol-set2}
\end{figure*}

\begin{deluxetable}{lcccccccccccc}
\tablecaption{Polarization properties obtained by fitting the \textsc{xspec} model \texttt{const*tbabs*polconst*comptt} to the IXPE $I$, $Q$, and $U$ data. \label{tab:xspec_IQU}}
\tablehead{
& & \colhead{$\chi^2$ / dof} & \colhead{PD [\%]} & \colhead{PA [deg]} 
}
\startdata
 P2 & & 594 / 589 &  $1.3\pm0.2$ &     $40\pm5$ \\
 P3 & & 603 / 589 &  $1.1\pm0.3$ &    $-10\pm7$ \\
\hline
HR1 & & 609 / 583 &  $1.1\pm0.3$ &     $43\pm9$ \\
HR2 & & 551 / 589 &  $1.1\pm0.2$ &     $27\pm6$ \\
HR3 & & 549 / 587 &  $1.6\pm0.4$ &    $-24\pm7$ \\
\enddata
\tablecomments{Uncertainties are at 68\% confidence level.}
\end{deluxetable}

\begin{deluxetable}{cccccc}
\tablecaption{Polarization for different phase intervals and for different hardness ratios using \texttt{pcube}. \label{tab:pol}}
\tablehead{
\colhead{} & \colhead{PD [\%]} & \colhead{PA [deg]} & \colhead{$Q/I$ [\%]} & \colhead{$U/I$ [\%]} & MDP [\%]
}
\startdata
        P1 & $1.1\pm1.4$ & $-2\pm34$ & $1.1\pm1.4$ & $-0.1\pm1.4$ & 4.1 \\
        P2 & $1.6\pm0.3$ & 37$\pm$5 & $0.4\pm0.3$ & $1.5\pm0.3$ & 0.79 \\
        P3 & $1.4\pm0.3$ & $-12\pm7$ & $1.2\pm0.3$ & $-0.5\pm0.3$ & 0.98 \\
 \hline 
HR1 & $1.6\pm0.4$ & $40\pm7$ &  $0.2\pm0.4$ & $1.5  \pm0.4$ & 1.2 \\
HR2 & $1.3\pm0.3$ & $24\pm6$ &  $0.9\pm0.3$ & $1.0  \pm0.3$ & 0.79 \\
HR3 & $1.9\pm0.5$ & $-26\pm7$ & $1.2\pm0.5$ & $-1.5 \pm0.5$ & 1.5 \\
\enddata
\tablecomments{Uncertainties are at 68\% confidence level. The Minimum Detectable Polarization (MDP) is the maximum polarization produced by statistical fluctuations at 99\% confidence level.}
\end{deluxetable}

\subsection{Polarization Along the Orbital Phase Intervals}

We first studied the polarization degree (PD) and angle (PA) into each single phase interval in the whole IXPE 2--8\,keV energy band: Figure~\ref{fig:pol-set2}--top shows a polar plot representing the PD and PA confidence regions in the three phase intervals defined in Figure~\ref{fig:rate-HR-trends}. The polarization in the first phase interval (P1) --- the low-hard state --- is unconstrained, as expected due to the low counts; the polarization in the remaining two phase intervals --- P2 and P3 --- is significantly detected at a confidence level (CL) higher than 99\%. We also observe a clear rotation of the PA by \angleseg\ between P2 and P3; a 90$\degr$ rotation is not consistent with this result at 99.5~\% CL. The numerical values of polarization obtained in this analysis are reported in Table~\ref{tab:pol}. The results obtained in these phase intervals using \textsc{xspec} are summarized in Table~\ref{tab:xspec_IQU}.

We also attempted to measure the polarization properties in different energy bands (bottom panels of Figure~\ref{fig:pol-set2}). 
No evidence for an energy dependence at 90\% CL is observed in any energy band; however, there are low significance hints of PD energy dependence in P2 and P3. There is also a low significance indication of a PA rotation with energy in P1: the PA at lower energies is consistent with the PA measured in P3, while at higher energies it is similar to that of P2.  

\begin{figure*}
\centering
\includegraphics[width=0.8\linewidth]{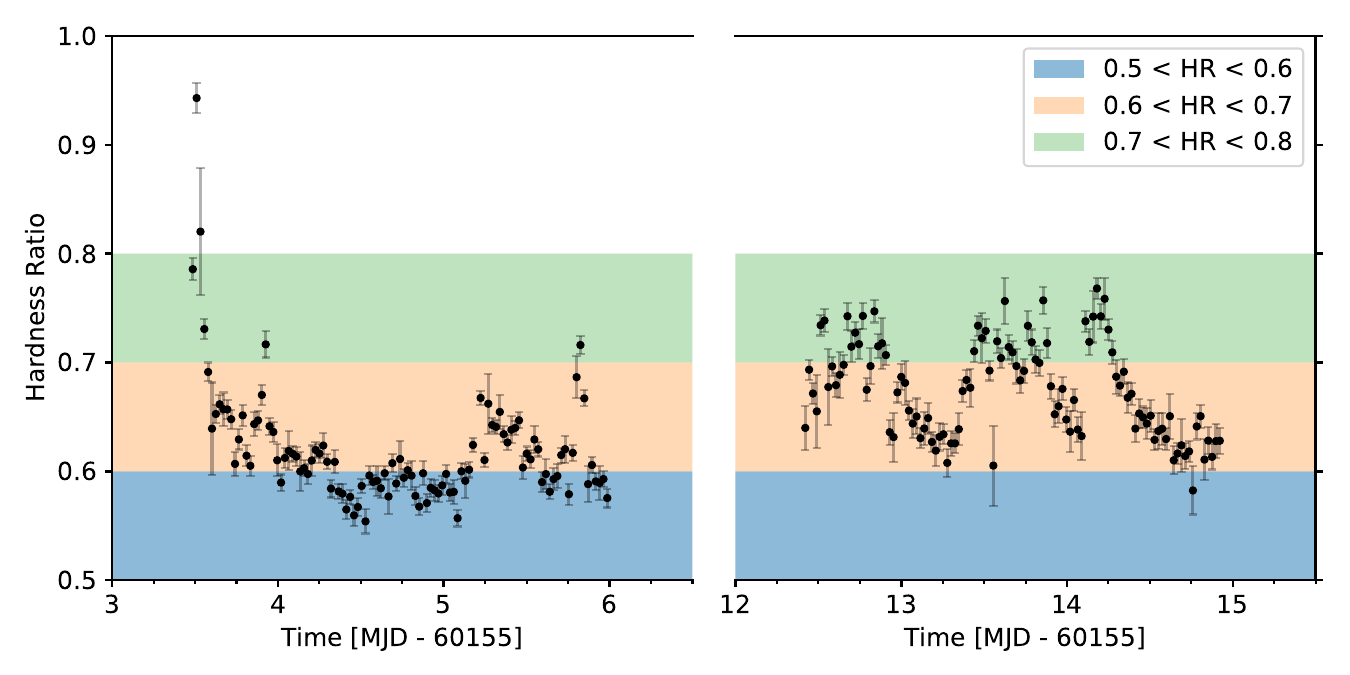}
\includegraphics[width=0.5\linewidth]{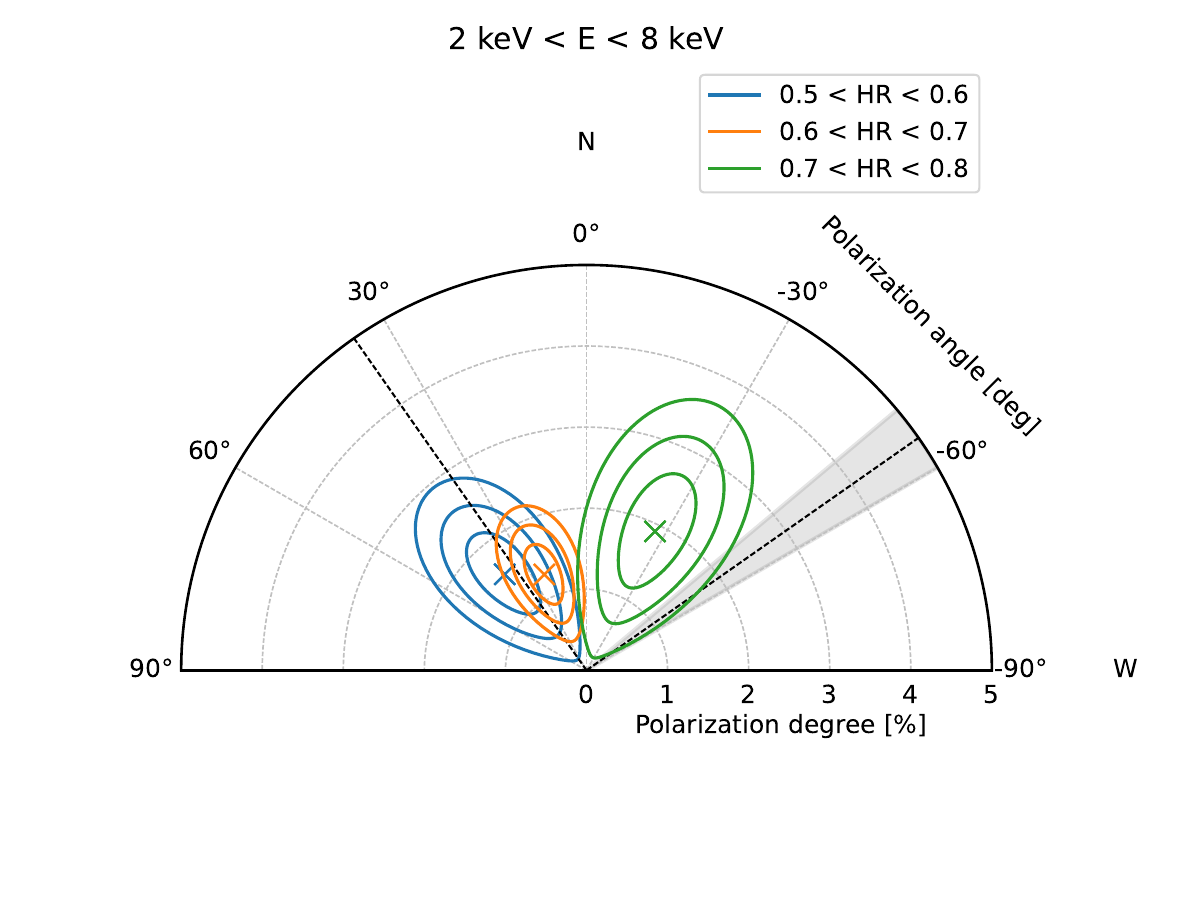}
\includegraphics[width=0.95\linewidth]{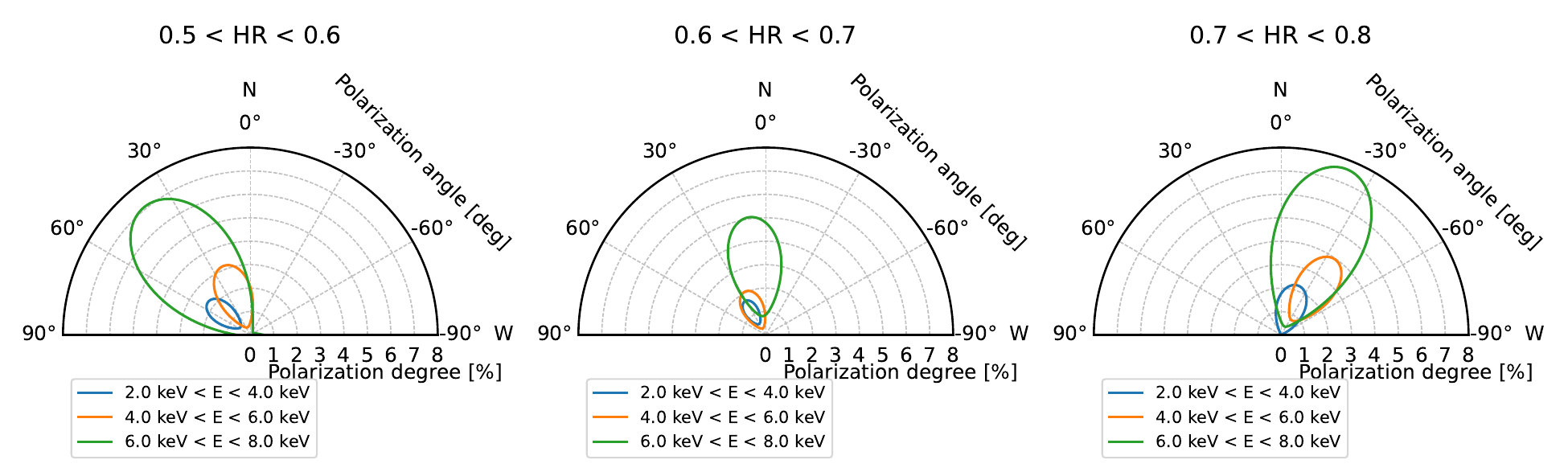}
\caption{Study of polarization binned in hardness ratio. (Top) HR trend along time (2000\,s time binning), the colored bands identify the chosen binning for HR. (Middle) Polar plot of polarization, computed using the model-independent \texttt{pcube} algorithm from the \textsc{ixpeobsssim} software \citep{2022SoftX..1901194B}, for events in different intervals of HR reported in the top panel, in the energy band 2--8 keV where the shaded region indicates the direction of the jet (see discussion), and the black lines indicate this direction and its orthogonal direction. Contours are reported at 68\%, 95\%, and 99\% confidence levels. (Bottom) Polarization in different energy band. Contours are reported at 90\% confidence level. 
}
\label{fig:pol-hr}
\end{figure*}

\subsection{Polarization as a function of the Hardness Ratio}

Since HR varies during the orbit --- even inside the phase intervals we considered above --- we studied the polarization in different HR states. This analysis is also useful to have an idea, given the difficulties in the determination of the spectral components, for the polarization of the different emission regions/components.
We used the HR values in the time bins of Figure~\ref{fig:rate-HR-trends}--bottom to define three different HR intervals: HR1 in the range 0.5--0.6, HR2 in the range 0.6--0.7 and HR3 in the range 0.7--0.8. Given the large uncertainty on each HR value --- at a level of $\sim$0.05 --- we used the average HR in larger time bins of 2000\,s each to populate the three HR intervals in the polarimetric analysis. This average curve is reported in Figure~\ref{fig:pol-hr}--top.

The polarization computed from events in these HR bins is shown in Figure~\ref{fig:pol-hr}; the numerical values are reported in Table~\ref{tab:pol}, while these obtained with \textsc{xspec} are in Table~\ref{tab:xspec_IQU}. The two estimates are compatible. In the 2--8 keV energy band the PD is compatible in the three HR intervals, while the PA shows a gradual rotation as the HR changes, with a total rotation of the PA by \anglehr\ between the lowest and the highest hardness-ratio bins; a 90$\degr$ rotation is not consistent with this result at 68.2\% CL. 

The polarization in different energy bands is reported in Figure~\ref{fig:pol-hr}--bottom. 
No evidence for an energy trend at 90\% CL is observed in any energy band, but there are only hints of an increase of PD with energy.

\section{Discussion and Conclusions}

\begin{figure*} 
\centering
\includegraphics[width=7cm]{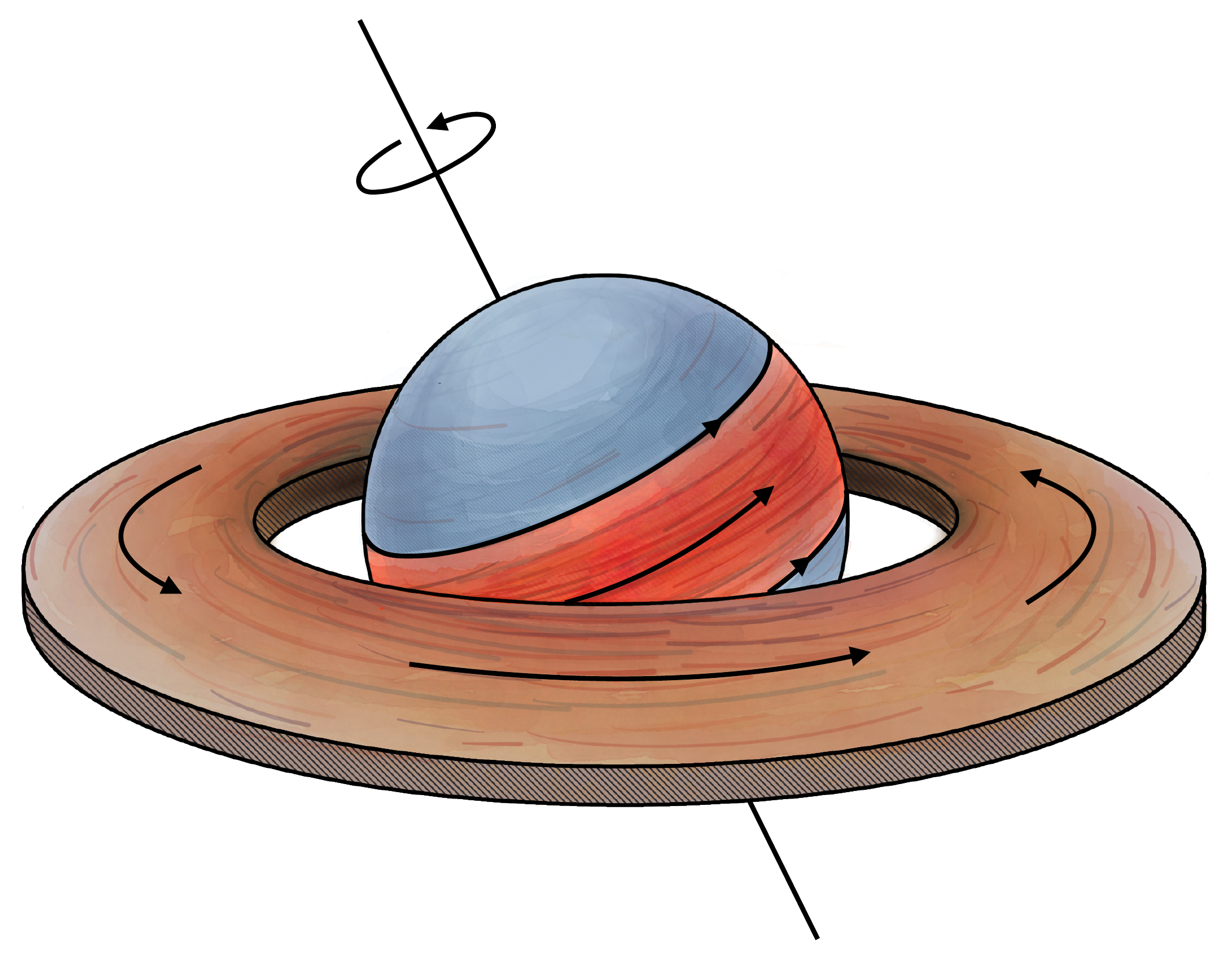}
\hspace{1.cm}
\includegraphics[width=7cm]{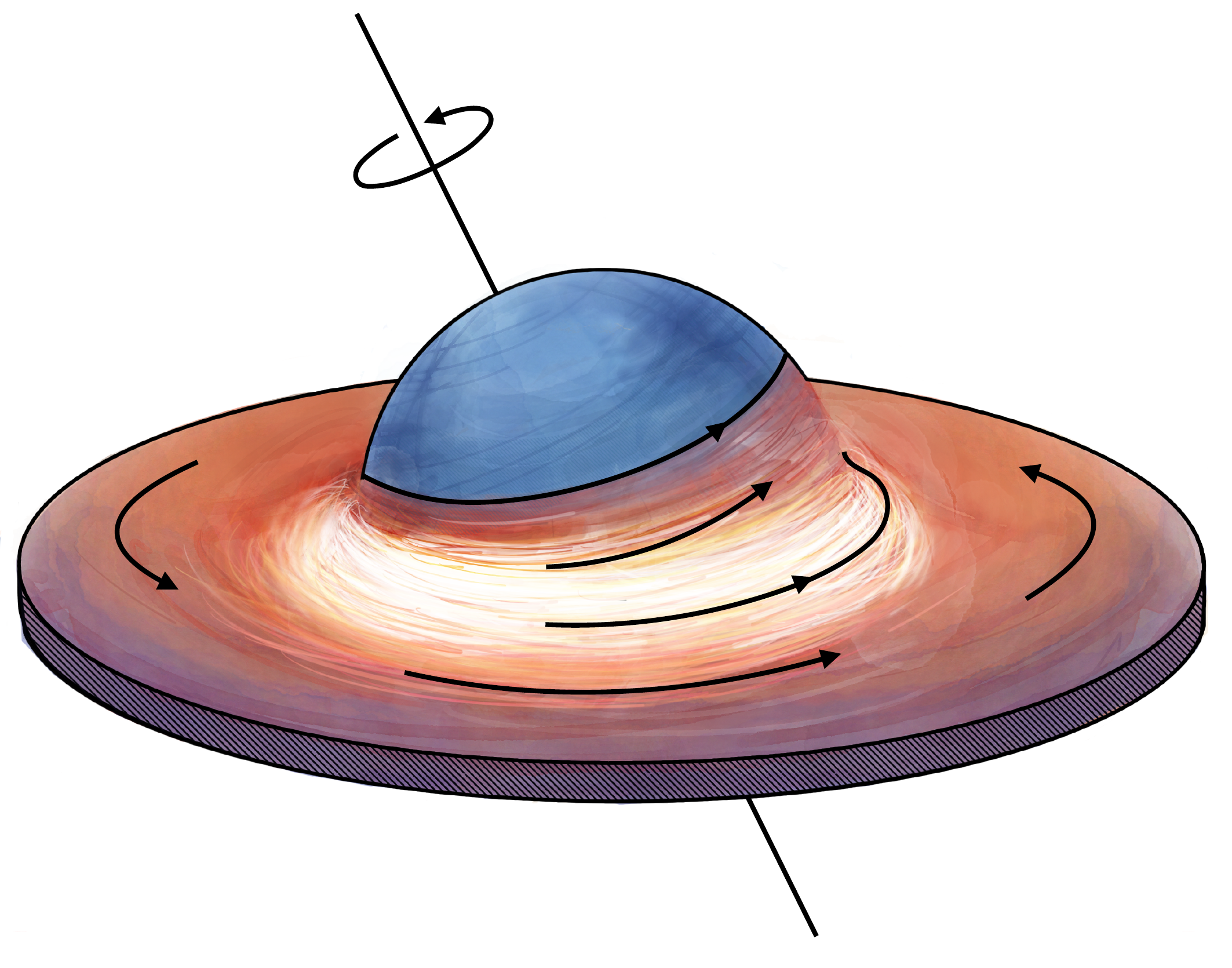}
\caption{
Illustration of a possible accretion geometry in \source. (Left) Low accretion-rate case, when there is a gap between the disk and the NS surface, and the full SL is developed. (Right) High accretion-rate case, where the disk touches the NS surface, and the BL is emitting (with a PA almost perpendicular to the symmetry axis of the disk).
}
\label{fig:illustration}
\end{figure*}

We studied for the first time the X-ray polarization of the NS-XRB \source. 
With the available statistics, there is no significant variation with energy (see lower panels of Figures~\ref{fig:pol-set2} and \ref{fig:pol-hr}). In other X-ray binaries observed by IXPE, an increase of the PD with energy was observed, but in this case we only have hints of such an increase; we also find low significance hints of a rotation of the PA with energy in the hard state at the beginning of the IXPE observation (P1 of Figure~\ref{fig:pol-set2}), which is the state where the Comptonization component is stronger. Comparing with \citet{2023ApJ...949L..10P} and \citet{2022MNRAS.514.2561G}, we find that the energy trend in polarization for \source observation is compatible with a shell or sandwich/wedge coronal geometry (with an inclination $<80\degr$), but not with a slab geometry --- for which a pronounced increase with energy would be expected. However, none of these two scenarios can fully explain our observation, where a change of the polarization is observed as a function of time and HR. This can be explained in terms of a scenario as the one of Figure~\ref{fig:illustration} which we present below, where a spreading layer and a boundary layer are present, but currently no such model is available in literature allowing for a deeper analysis capable to estimate the system inclination and the inclination of the NS axis with respect to the disk.

Along \source orbit, we observe a rotation of the PA by \angleseg\ between different phase intervals (Figure~\ref{fig:pol-set2}), while the PD stays constant within the same observing phase intervals. Being the accretion flow expected to be related to the orbital variations of \source, we performed an analogous study in HR intervals: we observe also in this case a constant PD with a rotation of the PA by \anglehr\  (Figure~\ref{fig:pol-hr}). 
The rotation between phase intervals is compatible with the rotation between HRs within 68\% CL.

From a spectral point of view, $\texttt{comptt}$ dominates in the IXPE energy band. However, at least two components are present, as reported in Section \ref{sec:spec}.
Thus the variations along the orbit can be due to a superimposition of two different components contributing in a different way along the phase intervals. Looking at the top panel of Figure~\ref{fig:pol-hr} we see that all the phase intervals are dominated by HR2 ($\sim60\%$), with a contribution from HR1 ($\sim35\%$) in P2, and from HR3 ($\sim35\%$) in P3. This confirms a scenario where the accretion flow --- and the HR --- varies along the orbit, with the harder state gradually becoming dominant as we are further away from the end of the dip (close to the beginning of the first IXPE observation).  
This is compatible with a model in which the accretion disk changes during the orbit due to its eccentricity \citep{1999MNRAS.308..415J}. In this model the modulation in the X-ray luminosity is due to orbital variations in the mass accretion of the compact star; during the periastron passage, the companion star overfills its Roche lobe, and the accretion disk is perturbed, through both tidal interaction and a sudden surge of material inflow, triggering an X-ray outburst. 

We can interpret variations of the PA as due to two spectral components with significantly different PAs: a lower energy one dominating at low HR, and a harder one dominates at higher HR. In the intermediate HRs the two components are mixed. 
Such a two-component model, composed e.g. of a multicolor blackbody from the accretion disk and a Comptonized component, is the obvious candidate to explain two components in the polarization, and so has been proposed for the other IXPE observations of weakly magnetized accreting NSs \citep{2023A&A...676A..20U, 2023ApJ...943..129C, 2023MNRAS.519.3681F, 2023A&A...674L..10C,  2023ApJ...953L..22D}. 
From simple geometrical considerations, and in the absence of relativistic effects, each component can be expected to be polarized either parallel or orthogonal to its symmetry axis. 
The PA of the optically thick accretion disk is expected to be perpendicular to the position angle of the rotation axis. 
Relativistic effects may cause a small decrease (for counterclockwise rotation) of the PA by 5\degr--10\degr\  \citep{Loktev2022}. 
The Comptonization component can be associated either with the BL (which is coplanar with the accretion disk), or with the SL at the NS surface.
The PA of the BL is likely nearly aligned with the PA of the accretion disk.  
In the absence of relativistic effects, the PA of the optically thick SL emission is parallel to the rotation axis. 
Due to aberration and Doppler boosting, the emission is expected to be dominated by the part of the SL moving towards the observer, breaking the symmetry and causing a decrease (also for counterclockwise rotation) of the PA by up to $\sim$20\degr--30\degr\ depending of the parameters (Bobrikova  et al., in prep.). 
If the SL is optically thin, the PA may rotate by an additional 90\degr\ \citep{Sunyaev1985,Viironen2004}. 
However, it is clear from this consideration that the difference in the PA by 50\degr -- 60\degr\ -- as found for \source\ -- is impossible to produce. 

The coexistence of two components with such a large difference in the PA may be explained if their symmetry axes are not aligned.
This can be related, for example, to a misalignment of the NS angular momentum with respect to the orbital axis, this way causing a shift of the symmetry axis of the Comptonization region (associated with the SL) with respect to the disk \citep{Abolmasov2020}.  
We note here that \source is not the only source for which such a misalignment might be present, but there are other hints: in \mbox{Cyg X-2} the PAs of the two components are 66\degr\ apart \citep[see Fig.\,7 in][]{2023MNRAS.519.3681F}, while in \mbox{XTE~J1701$-$462} and GX 5$-$1 the difference is $\sim$40\degr\  \citep{2023A&A...674L..10C,Fabiani23}.
Also, X-ray polarimetry provided evidence for a misalignment in the X-ray pulsar Her~X-1 \citep{Doroshenko2022}.
Such a misalignment is in fact more likely for \source, than for other NS systems, as the system is younger than 4600 years \citep{2013ApJ...779..171H}: if the newly-formed NS spins out of plane with respect to the binary system, there was not enough time to come to the alignment of the spinning axes.

At the same time, the accretion disk itself has a lower temperature compared to the Comptonization components \citep{Iaria2008}, and does not contribute significantly to the IXPE band. 
Thus the only other option for the second component is the BL. 
At low accretion rates (Figure~\ref{fig:illustration}--left), the disk is terminated at the innermost stable orbit of $\sim$3 Schwarzschild radii (i.e. about 13.5~km for a 1.5$M_\odot$ NS) which is likely larger than the NS radius of $\sim$12~km \citep[e.g.,][]{Nattila2017,Annala2022}. 
In this situation, the BL does not exist at all and matter freefalls on the NS surface, forming a SL. Thus, the PA would correspond to the orientation of the NS rotation axis on the sky. At high accretion rates (Figure~\ref{fig:illustration}--right), the thickness of the SL grows, connecting it to the accretion disk through the BL. In this case, the PA would be related to the symmetry axis of the disk.  

\citet{2023ApJ...958...52T} modeled \source as an accretion disk covered by a partially covering media, and interpreted the different phases of the orbit as changes in these two components. Their observations cover the dip for much longer than the IXPE polarimetric ones, where there is no significant polarimetric information for this phase; if we had observed the dip for longer with IXPE, we might have expected a high polarization degree due to obscuration, as observed in black hole systems \citep{2023arXiv230301174V, 2023MNRAS.519...50U}. It is interesting to note how their model is very good at predicting observational features such as lines, while to interpret the polarization we need a geometric model dealing with different features --- such as the model we outlined in this paper.

In order to understand the geometry of the inner accretion flow in \source, it is now worth relating the observed PAs to the orientation of the jet. \source is among the few NS-XRB showing jets both in the radio and the X-rays. 
The position angle of the (approaching) jet measured in the radio lies in the range 110\degr--140\degr\ \citep{1998ApJ...506L.121F,2008MNRAS.390..447T,2010ApJ...719L.194S,2012MNRAS.419L..54C,2012MNRAS.419L..49M}.
Also in the X-rays the signatures of the (receding) jet is found in the north-west direction at position angles of about $-70\degr$ and $-35\degr$ \citep{2007ApJ...663L..93H,2009MNRAS.397L...1S}, while the approaching jet is seen in the PA interval of 90\degr--150\degr\  \citep{2009MNRAS.397L...1S}. 
Thus the average jet direction seems to be nearly orthogonal to the direction of the X-ray polarization in P2 (and at H1 and H2; see Table~\ref{tab:pol}), which we associate with the BL emission. 
On the other hand, the X-ray polarization at the highest hardness ratio (HR3) is $\sim1.3 \sigma$ apart the the jet direction ($-35\degr$ or 140\degr--150\degr). 
Finally, the X-ray PA of $-12\degr$ during P3 is clearly neither parallel, nor perpendicular to the jet. 
Associating the observed PA with the SL implies a misalignment of the NS's angular momentum from the orbital axis by about 30\degr\  (Figure~\ref{fig:illustration}). 
Because the PA in this case is larger than the jet position angle,  the rotation of the SL (and of the disk) has to be clockwise, corresponding to an inclination exceeding 90\degr.

Although a large spread in the position angles measured for the jet can be explained by precession \citep{2012MNRAS.419L..54C,2010ApJ...719L.194S}, such an interpretation does not work for the variations of the X-ray PAs, because of much shorter time scales involved and detection of different PAs at different HRs. 
This gives further support to the interpretation that variations of the PA are caused by different spectral components (accretion disk, BL, and SL) dominating at different times. 
With the data at hand it is impossible to extract those components from the spectra; 
variations of the PA, however, strongly support the idea that the NS angular momentum is misaligned from the orbital one, which is a necessary requirement for the precession to operate.   
The X-ray polarimetric data, thus, provide a unique view of the geometry of the accreting NS \source.

\section*{Acknowledgments}

The Imaging X-ray Polarimetry Explorer (IXPE) is a joint US and Italian mission.  The US contribution is supported by the National Aeronautics and Space Administration (NASA) and led and managed by its Marshall Space Flight Center (MSFC), with industry partner Ball Aerospace (contract NNM15AA18C).  The Italian contribution is supported by the Italian Space Agency (Agenzia Spaziale Italiana, ASI) through contract ASI-OHBI-2022-13-I.0, agreements ASI-INAF-2022-19-HH.0 and ASI-INFN-2017.13-H0, and its Space Science Data Center (SSDC) with agreements ASI-INAF-2022-14-HH.0 and ASI-INFN 2021-43-HH.0, and by the Istituto Nazionale di Astrofisica (INAF) and the Istituto Nazionale di Fisica Nucleare (INFN) in Italy.  This research used data products provided by the IXPE Team (MSFC, SSDC, INAF, and INFN) and distributed with additional software tools by the High-Energy Astrophysics Science Archive Research Center (HEASARC), at NASA Goddard Space Flight Center (GSFC).
This research has made use of the MAXI data provided by RIKEN, JAXA, and the MAXI team. 

We acknowledge support from the Academy of Finland grants 333112,  355672, and  349144 (JP, AV, SST) and the German Academic Exchange Service (DAAD) travel grant 57525212 (VD). 
VKr acknowledges support from the Finnish Cultural Foundation.
FA and APap acknowledge financial support from the INAF Research Grant ``Uncovering the optical beat of the fastest magnetised neutron stars (FANS)'' and from the Italian Ministry of University and Research (MUR), PRIN 2020 (prot. 2020BRP57Z) ``Gravitational and Electromagnetic-wave Sources in the Universe with current and next-generation detectors (GEMS)''. 
This work is partially supported by National Key R\&D Program of China (grant No. 2023YFE0117200).
IL was supported by the NASA postdoctoral program at the Marshall Space Flight Center, administered by Oak Ridge Associated Universities under contract with NASA.
FX is supported by the National Natural Science Foundation of China (Grant No. 12373041).

\vspace{5mm}
\facilities{IXPE, NICER, NuSTAR, MAXI}

\software{\textsc{ixpeobssim} \citep{2022SoftX..1901194B}, \textsc{xspec} \citep{Arnaud1996}, \textsc{HEASoft} \citep{2014ascl.soft08004N}
}

\bibliography{CirX1_discovery_paper}{}
\bibliographystyle{aasjournal}

\end{document}